\documentclass[12pt, a4paper]{article}
\usepackage{latexsym,amsfonts,amssymb}
\usepackage{graphicx}
\usepackage{indentfirst}
\usepackage{cite}
\usepackage[cp1251]{inputenc}

\newtheorem{theorem}{Theorem}

\newtheorem{definition}{Definition}

\newtheorem{remark}{Remark}

\begin{document}

\begin{center}
 {\Large \bf Lie symmetries of  nonlinear boundary value problems}

\medskip

{\bf Roman Cherniha$^{\dag \ddag}$} {\bf and  Sergii Kovalenko$^\dag$}
\\
{\it $^\dag$~Institute of Mathematics, Ukrainian National Academy of Sciences,
\\
3 Tereshchenkivs'ka Street, Kyiv 01601, Ukraine}
\\
{\it $^{\ddag}$~Department of Mathematics, National University `Kyiv Mohyla Academy',
\\
2 Skovoroda Street, Kyiv 04070, Ukraine}

\medskip
 E-mail: cherniha@imath.kiev.ua and kovalenko@imath.kiev.ua
\end{center}

\begin{abstract}
Nonlinear boundary value problems (BVPs) by means of the classical
Lie symmetry method are studied. A new definition of Lie invariance
  for BVPs  is  proposed  by the  generalization
  of existing those  on much wider class of
  BVPs.
A class of  two-dimensional nonlinear boundary value problems,
modeling the process of melting and evaporation of metals, is
studied  in details. Using the definition proposed, all possible Lie
symmetries and the relevant reductions (with physical meaning) to
BVPs for ordinary differential equations  are constructed. An
example how to construct exact solution of the problem with
correctly-specified coefficients is presented and compared with the
results of numerical simulations published earlier.

\medskip

   2010 Mathematics Subject Classification : 22E70, 35K61, 80A22.
\end{abstract}

\section{\bf Introduction}

It is
well known that principle of linear superposition cannot be applied to
generate new exact solutions to  {\it nonlinear} partial
differential equations (PDEs).
 Thus, the classical methods (the Fourier method, the method of the Laplace
 transformations, and so forth) are not
applicable for solving  nonlinear PDEs. While there is no existing
general theory for integrating nonlinear PDEs,
 construction of  particular exact solutions for these equations
 is  a non-trivial and  important problem. Now the most popular methods for construction of  exact solutions
to  non-integrable  nonlinear PDEs  are  the Lie, Lie-B\"acklund and
conditional  symmetry methods  \cite{ovs, b-k, ibr-85, olv, fss,
bl-anco02}.  Although these   methods are very powerful provided the
relevant symmetry is known, several other approaches for solving
non-integrable  nonlinear  PDEs were independently suggested  during
the  last decades. Among them
 the method  of compatible differential constraints  \cite{yanenko-1984, olv-94},
the method  of linear invariant subspaces  \cite{gal-svi-07},
 $\tanh$-method  and its various modifications  \cite{malfield-96, kudrya-05, wazwaz-07, wang-li-zhang, ma-10}, the method of   additional  generating  conditions
  \cite{ch96, ch98-c}, and
  the transformed rational function method  \cite{ma-09} should be marked out (see, e.g., Supplements in  \cite{pol-za-2004} about   other methods).

The Lie symmetries are widely applied to study nonlinear
differential equations (including multi-component systems of PDEs)
since 60-s of the last century, notably, for constructing their
exact solutions. Nevertheless there are a huge number of papers and
many excellent books (see, e.g.,  \cite{ovs, b-k, olv, fss,
bl-anco02} and papers cited therein) devoted to such applications,
one may note that a very small number of  them involve Lie
symmetries to solve boundary value problems for the given  PDEs. To
the best of our knowledge, the first papers in this directions were
published in the beginning of 1970-s \cite{pukh-72} and
\cite{bl-1974} (the extended versions of these papers are presented
in books \cite{pukh-et-al-98} and \cite{b-k}, respectively). The
books, which highlight essential role of Lie symmetries in solving
boundary value problems (BVPs) and present several examples, were
published only in 1989 \cite{b-k,rog-ames-89}.

 The main object of this paper is a nonlinear BVP of Stefan type,
which belongs to the class of BVPs with free (moving) boundaries.
Boundary value problems  of Stefan type  are widely used  in
mathematical  modeling  a huge number of processes, which   arise in
physics,  biology  and  industry \cite{alex93, bri-03, crank84,
ready, rub71}. Nevertheless these processes can be very different
from formal point of view, they have  the common peculiarity,
unknown moving boundaries. Movement of unknown boundaries are
described by famous Stefan boundary conditions \cite{rub71, st}. It
is well-known that exact solutions of BVPs of Stefan type  can be
derived only in exceptional cases and the relevant list is rather
short  at the present time (see \cite{alex93, br-tr02, ch93,
ch-od90, brd96, voller04, voller10, barry08} and papers cited
therein).

Nevertheless BVPs with free  boundaries are   more complicated
objects than the standard BVPs with the fixed boundaries, it can be
noted that the Lie symmetry method should be more applicable just
for solving problems with moving boundaries. In fact, the structure
of such boundaries may depend  on invariant variable(s) and this
gives a possibility to reduce the given BVP to that of lover
dimensionality. This is the reason why different  authors applied
the Lie symmetry method to BVPs with free  boundaries ignoring BVPs
with fixed boundaries \cite{ben-olv-82, bl-1974, ch-kov-09, pukh-72,
pukh-06}.

The paper is organized as follows.
  In  Section 2, we discuss the existing definitions of Lie invariance
  for BVPs  and propose their generalization on much wider class of
  BVPs. As an example the direct  application of the definition  for the
  well-known BVP with the fixed boundaries is presented.  In  Section 3,
we apply the definition derived to the class of (1+1)--dimensional
BVPs of Stefan type used to describe melting and evaporation of
materials in the case when  their surface is exposed to a powerful
flux of energy \cite{ch-od90,ch-od91}. The  result obtained  is an
essential generalization of paper \cite{ch-kov-09}.
 In Section 4,   we  reduce the
problem   to BVPs for  ordinary differential equations, using Lie
symmetry operators obtained in the previous section. An example how
to construct exact solution of the problem with correctly-specified
coefficients is also presented. Finally, we present conclusions  in
the last section.

\section{\bf Definition of Lie invariance for BVPs}

We start from a definition of invariance of a BVP under the given
infinitesimal operator presented in \cite{bl-anco02, b-k} and
restrict ourselves on the case when the basic equation of BVP is an
(1+1)--dimensional evolution PDE of $k$th--order ($k\geq 2$). In
this case the relevant BVP may be formulated as follows:
\begin{equation}\label{ad-1}
u_t=F\left(x, u, u_x, \ldots, u_{x}^{(k)}\right), \ (t,x) \in \Omega
\subset \mathbb{R}^2
\end{equation}
\begin{equation}\label{ad-2}
s_a(t,x)=0: \ B_a \left(t,x, u, u_x, \ldots, u_{x}^{(k-1)}\right) =
0, \ a = 1, 2, \ldots, p,
\end{equation}
where $F$ and  $B_a$ are smooth functions in the corresponding
domains,   $\Omega$  and $s_a(t,x)$ are  a domain with smooth
boundaries and smooth curves, respectively. Hereafter the subscripts
$t$ and $x$ denote differentiation with respect to these variables,
$u_{x}^{(j)} = \frac{\partial^j u}{\partial x^j}, j = 1,2,
\ldots,k$. We assume that BVP (\ref{ad-1}) and (\ref{ad-2}) has a
classical solution (in a usual sense).

 Consider the infinitesimal generator
\begin{equation}\label{ad-3}
X = \xi^0 (t,x)\frac{\partial}{\partial t}+\xi^1
(t,x)\frac{\partial}{\partial x} + \eta
(t,x,u)\frac{\partial}{\partial u},
\end{equation}
(hereafter  $\xi^0, \xi^1 $ and $\eta$ are known smooth functions),
which defines a Lie symmetry acting on both $(t,x,u)$--space as well
as on its projection to $(t,x)$--space. Let $X^{(k)}$ be the
$k$th--prolongation of the generator $X$ calculated by the
well-known prolongation formulae (see, e.g. \cite{olv, ovs}).

\begin{definition}
\cite{b-k} The Lie symmetry $X$ of the form
(\ref{ad-3}) is admitted by the boundary value problem
(\ref{ad-1}) and (\ref{ad-2}) if and only if:
\begin{itemize}
\item[(a)] $ X^{(k)} \left(F\left(x, u, u_x, \ldots , u_{x}^{(k)}\right)
-u_t \right)=  0 $ when $u$ satisfies (\ref{ad-1});
\item[(b)] $X (s_a(t,x)) = 0$ when $s_a(t,x) = 0, \ a = 1,2, \ldots,p$;
\item[(c)] $ X^{(k-1)} \left(B_a \left(t,x, u, u_x, \ldots ,
u_{x}^{(k-1)}\right) \right) =  0 $ when $B_a \vert_{ s_a(t,x) = 0}=
0$, \ $a = 1,2, \ldots,p$.
\end{itemize}
\end{definition}

 The definition can  straightforwardly be extended on BVPs
 for a system of PDEs.  However, one easily  notes
 that this definition  cannot  be applied to BVPs
with free boundaries, because such problems contain  moving
surfaces, say $S_b(t,x)=0, \ b = 1, \ldots, q,$ where  $S_b(t,x)$
are unknown functions. Obviously, these functions should be
interpreted as additional variables. In \cite{ben-olv-82} (see
Appendix 2), a criteria of invariance  for  BVP with a free boundary
was formulated. Another deficiency of Definition 1 appears if one
consider BVPs  in the unbounded domain $\Omega$ when the boundary
conditions for $ x = \infty $ arise. In fact, item (b) has no sense
in this case and cannot be replaced by the natural  passage to the
limit, i.e., $ x = L, \ L \to \infty .$ Probably this  deficiency
for the first time was noted in \cite{king-91} (see Section 4.3)
where the transformation $ x = 1/y$ was suggested to avoid the
non-regular manifold generated by $ x = \infty $.

  Now we present a definition which  takes into account
  all possible boundary conditions and is applicable to a wide  range of BVPs.
Consider a BVP for a  system of $n$ evolution equations ($n \geq 2$)
with $2$ independent $(t, x)$ and $n$ dependent $u = (u_1, u_2,
\ldots, u_n)$ variables. Let us assume that the $k$th--order ($k\geq
2$) basic equations of evolution type
\begin{equation}\label{1}
u_t^i=F^i \left(x, u, u_x, \ldots , u_{x}^{(k)}\right), \ i = 1,
\ldots, n
\end{equation}
\noindent are  defined on a domain ${\Omega} \subset \mathbb{R}^2 $
with smooth boundaries. Consider  three types of boundary and
initial conditions, which can arise in applications:
\begin{equation}\label{2}
s_a(t,x)=0: \ B^{j}_a \left(t,x, u, u_x, \ldots ,
u_{x}^{(k_{a}^j)}\right) = 0,\ a = 1, \ldots, p, \, j =1,\ldots,n_a,
\end{equation}
\begin{equation}\label{3}
S_b(t,x)=0: \ B^{l}_b \left(t,x, u, \ldots , u_{x}^{(k_b^l)}, S_b,
\frac{\partial S_b}{\partial t}, \frac{\partial S_b}{\partial
x}\right) = 0, \ b = 1, \ldots, q, \, l =1,\ldots,n_b,
\end{equation}
\noindent and
\begin{equation}\label{4}
\gamma_c(t,x)=\infty: \ \Gamma^{m}_c \left(t,x, u, u_x, \ldots ,
u_{x}^{(k_{c}^m)}\right) = 0, \ c = 1, \ldots, r, \, m = 1, \ldots,
n_c.
\end{equation}
\noindent Here $k_{a}^j < k, \ k_{b}^l< k$ and  $k_{c}^m < k$ are
the given numbers, $s_a(t,x)$ and $\gamma_c(t,x)$ are the known
functions, while the functions $S_b(t,x)$ defining free boundary
surfaces must be found. We assume that all functions arising in
(\ref{1})--(\ref{4}) are sufficiently smooth so that a classical
solution exists for this BVP.

Consider an $N$--parameter  (local) Lie group  $G_N$ of point
transformations of variables $(t,x,u)$ in the Euclidean space
$\mathbb{R}^{n+2}$ (open subset of $\mathbb{R}^{n+2}$), which is
given by equations
\begin{equation}\label{5}
t^{\ast} = T(t,x,\varepsilon), \ \ x^{\ast} = X(t,x,\varepsilon), \
\  u^{\ast}_i = U_i(t,x,u,\varepsilon), \ i = 1, \ldots, n,
\end{equation}
\noindent where $\varepsilon = (\varepsilon_1, \varepsilon_2,
\ldots, \varepsilon_N)$ are the group parameters. According to the
general Lie group theory, one may construct the  corresponding
$N$--dimensional Lie algebra $L_N$ with the basic  generators
\begin{equation}\label{6}
X_\alpha = \xi^0_\alpha \frac{\partial}{\partial t}+\xi^1_\alpha
\frac{\partial}{\partial x} + \eta^1_\alpha \frac{\partial}{\partial
u^1}+ \ldots +\eta^n_\alpha \frac{\partial}{\partial u^n}, \ \alpha
= 1,2, \ldots, N,
\end{equation}
\noindent where $\xi^0_\alpha = \left. \frac{\partial
T(t,x,\varepsilon)}{\partial \varepsilon_\alpha}\right
\vert_{\varepsilon = 0}, \ \xi^1_\alpha = \left. \frac{\partial
X(t,x,\varepsilon)}{\partial \varepsilon_\alpha}\right
\vert_{\varepsilon = 0}, \ \eta^i_\alpha = \left. \frac{\partial
U_i(t,x,u,\varepsilon)}{\partial \varepsilon_\alpha}\right
\vert_{\varepsilon = 0}$.

Consider the Lie algebra $L_N$ in the extended space
$\mathbb{R}^{n+q+2}$ of the variables $(t,x,u,S)$, where $S = (S_1,
..., S_q)$ are new dependent variables with respect to $t$ and $x$.
In the extended space $\mathbb{R}^{n+q+2}$,   the  Lie group
$\widetilde{G}_N$ corresponding to this algebra is given by
transformations
\begin{equation}\label{7}
t^{\ast} = T(t,x,\varepsilon), x^{\ast} = X(t,x,\varepsilon),
u^{\ast}_i = U_i(t,x,u,\varepsilon), S^{\ast}_b = S_b(t,x), \ i = 1,
\ldots, n, b = 1, \ldots, q.
\end{equation}
Now we propose a new  definition, which is based on  the standard
definition of differential equation invariance as an invariant
manifold ${\cal{M}}$ in the relevant space of variables and on the
prolongation theory \cite{olv}.

\begin{definition}
A boundary value problem (\ref{1})--(\ref{4}) is called to be invariant with respect to the Lie group $\widetilde{G}_N$ (\ref{7}) if:
\begin{itemize}
\item[(a)] the manifold determined by Eqs. (\ref{1}) in the space of variables $\left(t,x,u, \ldots, u^{(k)}\right)$
 is invariant with respect to the
$k$th--order prolongation of the group $G_N$;
\item[(b)]each  manifold determined by conditions (\ref{2}) with
  any fixed number $a$  is invariant with respect
to the $k_a$th--order prolongation of the group $G_N$ in the space
of variables $\left(t,x,u, \ldots, u^{(k_a)}\right)$ , where $k_a =
\max \{k_{a}^j, \ j = 1, \ldots, n_a \}$;
\item[(c)] each  manifold determined by conditions (\ref{3}) with
  any fixed number $b$  is invariant with respect
to the $k_b$th--order prolongation of the group $\widetilde{G}_N$ in
the space of variables $\left(t,x,u, \ldots, u^{(k_b)}, S_b,
\frac{\partial S_b}{\partial t}, \frac{\partial S_b}{\partial
x}\right)$ ,  where $k_b=\max \{k_{b}^l, \ l =1,\ldots,n_b \}$;
\item[(d)]each  manifold determined by conditions (\ref{4}) with
  any fixed number $c$  is invariant with respect
to the $k_c$th--order prolongation of the group $G_N$ in the space
of variables $\left(t,x,u, \ldots, u^{(k_c)}\right)$ ,
 where $k_c = \max \{k_{c}^m, \ m = 1, \ldots, n_c \}$.
\end{itemize}
\end{definition}

\begin{definition}
The functions $u_i = \Phi_i(t,x), \,i = 1,
\ldots, n$ and $S_b = \Psi_b(t,x),\, b = 1, \ldots, q$  form  an
invariant solution of BVP (\ref{1})--(\ref{4}) corresponding to the
Lie group (\ref{7}) if:
\begin{itemize}
\item[(i)] the functions $u_i = \Phi_i(t,x)$ and $S_b = \Psi_b(t,x)$
 satisfy equations  (\ref{1})--(\ref{4});
\item[(ii)] the manifold  ${\cal{M}} = \{u_i = \Phi_i(t,x), \ i = 1, \ldots, n; \,
S_b = \Psi_b(t,x), \ b = 1, \ldots, q \}$ is an invariant manifold
of the Lie group (\ref{7}).
\end{itemize}
\end{definition}

\begin{remark}
Definition 2 can be   generalized on more
general systems (including hyperbolic and elliptic those) and
boundary conditions containing   high-order derivatives for
$S_b(t,x)$.
\end{remark}

\begin{remark}
If free boundaries are given in the  form $x = S_b(t)$, where $b = 1, \ldots, q$ then we simply take $S_b(t,x) \equiv x - S_b(t) = 0$. On the other hand, one  can formulate a
definition of Lie invariance for BVPs with  such form of the free
boundaries (see, e.g., \cite{ben-olv-82}). However, the form used in
Definition 2 is  more convenient for generalization on
multidimensional BVPs.
\end{remark}

Now we present  a non-trivial result  to illustrate Definition 2.
Let us consider the nonlinear  BVP modeling  the heat transfer in
semi-infinite solid rod assuming that  thermal diffusivity depends
on temperature and the rod is exposed to a periodical flux of energy
at the left endpoint. It should be noted that we neglect the initial
distribution of the temperature in the rod, i.e.,  consider the
process on the stage when the heat transfer already started. Thus
the nonlinear BVP reads as
\begin{eqnarray}
& & \frac{\partial u}{\partial t} = \frac{\partial}{\partial
x}\left(d(u)\frac{\partial u}{\partial x}\right), \ t>0, \ 0<x<+ \infty, \label{8} \\
& & \quad x = 0: d(u)\frac{\partial u}{\partial x} = q_0 \cos(\gamma t), \ t > 0, \label{10} \\
& & \quad x = + \infty: u = 0, \ t > 0, \label{11}
\end{eqnarray}
where $u(t,x)$ is an unknown temperature field, $d(u)$ is a thermal
diffusivity  coefficient, $q_0 \cos(\gamma t)$ is an energy flux. We
assume that all functions arising in (\ref{8})--(\ref{11}) are
sufficiently smooth, so that a classical solution exists for this
BVP.

Here  we restrict ourselves to the case  when the  thermal
diffusivity coefficient depends on the temperature as  a power low,
i.e. $d(u) = u^k$, where $k \in \mathbb{R}, \ k \neq 0$ (in the case
$k = 0$, the problem  is liner and can be solved by classical
methods, see, e.g., \cite{cr-jg59}). Notably, equation (\ref{8})
with $d(u) = u^k$ presents the most interesting cases of Lie
symmetry invariance \cite{ovs}. In the case $d(u) = u^k$ ($k \in
\mathbb{R}, $ $k \neq -\frac{4}{3}$), it admits  a four-dimensional
Lie group. The corresponding algebra $L_4$ possesses  the basic
operators $\langle
\partial_{t},
\partial_{x}, 2t \partial_{t} + x
\partial_{x},k x \partial_{x} + 2 u \partial_{u} \rangle$.
These operators generate the one-parameter Lie groups
\begin{eqnarray}
 & & T_1: \ t^{\ast} = t + \varepsilon_1, \ \  x^{\ast} = x, \ \ u^{\ast}
= u, \label{11a} \\ & & T_2: \ t^{\ast} = t, \ \ x^{\ast} = x +
\varepsilon_2, \ \ u^{\ast} = u, \label{11b} \\ & & T_3: \ t^{\ast}
= t e^{2 \varepsilon_3}, \ \ x^{\ast} = x e^{\varepsilon_3}, \ \
u^{\ast} = u, \label{11c}\\ & & T_4: \ t^{\ast} = t, \ \ x^{\ast} =
x e^{k \varepsilon_4}, \ \ u^{\ast} = u e^{2
\varepsilon_4}\label{11d},
\end{eqnarray}
respectively (hereafter  $\varepsilon_i, i=1,\dots,4$ are arbitrary
group parameters). If $k = - \frac{4}{3}$, then the additional
conformal generator $\langle
 x^2 \partial_{x} - 3xu\partial_{u} \rangle$ occurs, which  extends
$L_4$  to the five-dimensional Lie algebra $L_5$.
 Thus, the case $k = - \frac{4}{3}$
 should be examined separately.

\begin{table}
\caption{Lie invariance of BVP (\ref{8})--(\ref{11})}
\renewcommand{\arraystretch}{1.5}
\begin{center}
\tabcolsep=10pt
\begin{tabular}{ccccl}
 \hline\hline
  {no} & {$k$} & {$\gamma$} & {$q_0$} & {Lie groups  of invariance }\\
  \hline
  1. & $\forall$ & $\forall$ & 0 & $t^{\ast} = t e^{2 \varepsilon_2} + \varepsilon_1, \ x^{\ast} = x e^{\varepsilon_2 + k \varepsilon_3}, \ u^{\ast} = u e^{2 \varepsilon_3}$ \\
  2. & $\forall$ & 0 & $\forall, (q_0 \neq 0)$ & $t^{\ast} = t e^{(k + 2) \varepsilon_2} + \varepsilon_1, \ x^{\ast} = x e^{(k + 1) \varepsilon_2}, \ u^{\ast} = u e^{\varepsilon_2}$ \\
  3. & -2 & $\forall, (\gamma \neq 0)$ & $\forall, (q_0 \neq 0)$ & $t^{\ast} = t, \ x^{\ast} = x e^{- \varepsilon_1}, \ u^{\ast} = u e^{\varepsilon_1}$ \\
  \hline\hline
\end{tabular}
\end{center}
\end{table}

\begin{theorem}
All possible Lie groups  of invariance  of the nonlinear  BVP
(\ref{8})--(\ref{11}) with $d(u) = u^k,  \, k \neq 0$ for any
constants $q_0$ and $\gamma$ are presented in Table 1.
\end{theorem}

\noindent \textbf{Proof}. On the first step of the proof we will
consider BVP (\ref{8})--(\ref{11}) with the constant energy flux
$q(t) = q_0$, i.e. $\gamma = 0$. Let us study the case of arbitrary
power $k \neq -\frac{4}{3}$. First of all, we consider the
one-parameter Lie groups (\ref{11a})--(\ref{11d}) generated by the
basic operators of $L_4$.
 One easily notes  that  BVP (\ref{8})--(\ref{11}) is invariant with respect
 to the Lie group $T_1$ and isn't invariant under the Lie group $T_2$ since the boundary curve $x = 0$ isn't
 invariant with respect to the transformations (\ref{11b}).

According to item (b) of Definition 2, the boundary condition
(\ref{10}) is invariant with respect to the one-parameter group
$T_3$, if the manifold $\cal M$ = $ \{x = 0, \ u^k \frac{\partial
u}{\partial x} - q_0 = 0\}$ satisfies the conditions
\begin{equation}\label{dod1}
\left. x^{\ast} \right \vert_{\cal M} = 0, \ \left. (u^{\ast})^k
\frac{\partial u^{\ast}}{\partial x^{\ast}} - q_0 \right \vert_{\cal
M}= 0
\end{equation}
The first equation of  (\ref{dod1}) is an identity, while the second
equation leads to the expression $ q_0 e^{- \varepsilon_3} = q_0$,
 which immediately gives
\begin{equation}\label{16a}
q_0 \equiv 0.
\end{equation}
 The invariance of  (\ref{11}) under the one-parameter group $T_3$ is
obvious. Thus, BVP (\ref{8})--(\ref{11}) is invariant with respect
to the Lie group $T_3$ if and only if the restriction (\ref{16a})
takes place.

Dealing  in a similar way with the Lie group $T_4$, we obtain that
BVP (\ref{8})--(\ref{11}) is invariant with respect to $T_4$ only in
two cases:  $q_0 \neq 0$, $ k = -2$  and  $q_0 = 0$, $k \in
\mathbb{R}$. Indeed, according to item (b) of Definition 2, the
boundary condition (\ref{10}) is invariant with respect to $T_4$, if
conditions (\ref{dod1}) are satisfied on the manifold $\cal M$.
 Hence, we arrive at the restriction
\begin{equation}\label{dod2}
q_0 e^{(k+2) \varepsilon_4} = q_0,
\end{equation}
which  immediately  leads to  $k = -2$ provided $q_0 \not= 0$, and
$k \in \mathbb{R}$ if $q_0 = 0$. The invariance of (\ref{11}) under
the one-parameter Lie group $T_4$ is evident.

 Taking into account the restrictions considered above on $q_0$ and $k$,
 one concludes that BVP (\ref{8})--(\ref{11}) is invariant with respect
 to the two-parameter Lie group $T_1 \circ
T_4$ iff $k = -2, q_0 \neq 0$ and  with respect to the
three-parameter Lie group $T_1 \circ T_3 \circ T_4$ iff $k \in
\mathbb{R}, q_0 = 0$ (it is exactly case 1 of Table 1).

 To find other Lie groups of invariance, one needs  to consider
 a linear combination
of the basic operators of  $L_4$  excepting the operator
$\partial_t$(we remind that  the BVP  is  invariant  under  the Lie
group $T_1$ for arbitrary $q_0$ and $k$)
\begin{equation}\label{21}
X = 2 \lambda_3 t \partial_t + (\lambda_2 + (\lambda_3 + k
\lambda_4) x) \partial_x + 2 \lambda_4 u \partial_u,
\end{equation}
\noindent where $\lambda_i, \ i = 2, \ldots, 4$ are arbitrary
parameters and at least two of them are non-zero. If $\lambda_3=0$,
then  one arrives only at the results obtained above for the Lie
group $T_4$, if  $\lambda_4=0$  then the result obtained above for
the Lie group $T_3$ is recovered. If $\lambda_3\lambda_4 \neq 0$
then two possibilities occur: $\lambda_3 + k \lambda_4 \neq 0$ and
$\lambda_3 + k \lambda_4 = 0$. Consider the case $\lambda_3 + k
\lambda_4 \neq 0$ when operator (\ref{21}) generates the Lie group
\begin{equation}\label{dod3}
T_a: \ t^{\ast} = t e^{2 \lambda_3 \varepsilon_a}, \ x^{\ast} = x
e^{(\lambda_3 + k \lambda_4) \varepsilon_a} +
\frac{\lambda_2}{\lambda_3 + k \lambda_4} \left(e^{(\lambda_3 + k
\lambda_4) \varepsilon_a} - 1 \right), \ u^{\ast} = u e^{2 \lambda_4
\varepsilon_a}.
\end{equation}

Clearly, the boundary condition (\ref{11}) is invariant with respect
to $T_a$. Boundary conditions (\ref{10}) is invariant under $T_a$,
if and only if  conditions (\ref{dod1}) are satisfied. Now we
realize that the first equation of (\ref{dod1}) leads to the
requirement $\lambda_2 = 0$ while the second equation of
(\ref{dod1}) gives
\begin{equation}\label{dod6}
q_0 e^{((k + 2) \lambda_4 -\lambda_3) \varepsilon_a} = q_0
\end{equation}
Since $\lambda_3\lambda_4 \neq 0$, one immediately obtains
$\frac{\lambda_3}{\lambda_4} = k + 2$ provided $q_0 \neq 0$  and $k
\neq -2$. If $q_0 = 0$ then we immediately arrive at case 1 from
Table 1. On the other hand, the Lie group $T_1 \circ T_a$ transforms
into the group $T_1 \circ T_4$, when $k = -2$. Thus, we can conclude
that the BVP under study is invariant with respect to the
two-parameter Lie group $T_1 \circ T_a$ if and only if
\begin{equation}\label{k1}
q_0 \neq 0, \ \ \frac{\lambda_3}{\lambda_4} = k + 2.
\end{equation}
It is exactly case 2 of Table 1. The examination of the  case
$\lambda_3 + k \lambda_4 = 0$ leads to  case 2  with   $k=-1$. Thus,
the invariance of  BVP (\ref{8})--(\ref{11}) with $k \neq
-\frac{4}{3}$ is completely examined.

Now  we examine  the special case $k = -\frac{4}{3}$. One easily
checks that the one-parameter groups  (with $k = -\frac{4}{3}$)
listed in cases 1--2 of Table 1 are the groups of invariance of  BVP
(\ref{8})--(\ref{11}) (with $k = -\frac{4}{3}$) under the same
restrictions on  the constant $q_0$.

Thus, we need to examine whether   the BVP in question  can be
invariant with respect to a Lie group corresponding to any liner
combination of the basic operators of  $L_5$
\begin{equation}\label{23}
X = 2 \lambda_3 t \partial_t + (\lambda_2 + (\lambda_3 + k
\lambda_4) x + \lambda_5 x^2) \partial_x + (2 \lambda_4 - 3
\lambda_5 x) u \partial_u, \ \ \lambda_5 \neq 0.
\end{equation}
 \noindent To avoid cumbersome formulae, we  consider the
one-parameter Lie group corresponding to the pure conformal operator
\begin{equation}\label{22}
T_5: \ t^{\ast} = t, \ \ x^{\ast} = \frac{x}{1 - \varepsilon_5 x}, \
\ u^{\ast} = (1 - \varepsilon_5 x)^3 u.
\end{equation}
Let us study the invariance of the boundary condition (\ref{11}).
According to item (d) of  Definition 2, the following equalities
should take place
\begin{equation}\label{22a}
x^{\ast}\vert_{\cal N} = + \infty, \ \ u^{\ast}\vert_{\cal N} = 0,
\end{equation}
\noindent where $\cal N$ = $ \{x = + \infty, \ u = 0 \}$. However,
$x^{\ast}\vert_{\cal N} = - \frac{1}{\varepsilon_5}$. Thus, the
contradiction is obtained and we conclude that BVP
(\ref{8})--(\ref{11}) with $k = -\frac{4}{3}$ isn't invariant under
$T_5$.

In a quite similar way, one may show that the boundary condition
(\ref{11}) isn't invariant under any Lie group corresponding to
operator (\ref{23}).

Finally, to complete the proof, we must consider the  case, when the
flux of energy has periodical form, i.e.  $\gamma \neq 0$.
Obviously,  $q_0$ must be nonzero, otherwise we obtain the case
examined  above.  Since  calculations are quite  similar to the case
$\gamma = 0$  (an analog of formula (\ref{dod2}) plays a crucial
role to derive the special power $k=-2$),  we present  the result:
BVP (\ref{8})--(\ref{11}) with the periodic  energy flux $q(t) = q_0
\cos(\gamma t)$ is invariant  only with respect to the one-parameter
Lie group $T_4$ with $k = -2$ (case 3 from Table 1).

The proof is now completed. $\blacksquare$

\begin{remark}
Theorem 1 highlights that Definition 2 is
non-trivial because the power $k=-2$ isn't a special one for  Lie
invariance of  standard nonlinear heat equation (\ref{8}) with $d(u)
= u^k$, however, $k=-2$ is the  special power if one looks  for
Lie invariance of BVP (\ref{8})--(\ref{11}).
\end{remark}

\section{\bf Lie invariance of a class of (1+1)--dimensional nonlinear BVPs of Stefan type}

In this section we consider a class of (1+1)--dimensional  BVPs of
Stefan type used to describe melting and evaporation of materials in
the case that their surface is exposed to a powerful flux of energy.
Such  problems also  arise in   mathematical modeling of other
processes in biology (tumor growth) and physics (crystal growth).
The class of BVPs after some simplifications (like using the Goodman
substitution to transform  the basic   equations to the standard
heat  equations) can be written as follows
\begin{eqnarray}
& & \frac{\partial u}{\partial t}  =  \frac{\partial}{\partial
x}\left(d_{1}(u) \frac{\partial
u}{\partial x}\right),\label{34}  \\
 & & \frac{\partial v}{\partial t}  =  \frac{\partial}{\partial
x}\left(d_{2}(v)
\frac{\partial v}{\partial x}\right),\label{35} \\
& & \qquad S_{1}(t,x) = 0: \ d_{1}(u) \frac{\partial u}{\partial x}
= H_1(u)
V_1- q(t,u), \ V_1 = h(t,u),\label{36} \\
& & \qquad S_{2}(t,x) = 0: \ d_{2}(v_m) \frac{\partial v}{\partial
x} = d_{1}(u_m) \frac{\partial u}{\partial x} + H_2(v_m)
V_2, \ u = u_{m},  v =  v_{m},\label{37}
\\ & & \qquad x  =  +\infty: \ v = v_{\infty},\label{38}
\end{eqnarray}
where $u(t, x)$ and $v(t, x)$ are the unknown temperature fields;
$S_{k}(t,x),\  k=1,2$ are the unknown functions, which determine the
phase division boundaries  (they can be also presented in the form
$S_{k}(t,x) = x - s_k(t)$);
 $V_k(t,x) = - \frac{\partial
S_k}{\partial t} / \frac{\partial S_k}{\partial x}\  k=1,2$ are the
phase division boundary velocities; $q(t,u)$ is the known strictly
positive function presenting the energy flux being absorbed by the
material; $h(t,u)$ is the known non-negative  function  describing
dynamics of evaporation process; $ H_k, \ k=1,2$ are the known
strictly positive function presenting specific heat values per unit
volume of liquid and solid phases. The parameters $u_m, v_m$ and $
v_{\infty}$ are assumed to known, moreover, $ v_m \neq v_{\infty}$.

Here Eqs. (\ref{34}) and (\ref{35}) describe the heat transfer
process in liquid and solid phases, the boundary conditions
(\ref{36}) present evaporation dynamics on the surface $S_{1}(t,x) =
0$, and the boundary conditions (\ref{37}) are the famous  Stefan
conditions on the surface $S_{2}(t,x) = 0$ dividing the liquid and
solid phases. Assuming that the liquid  phase thickness is
considerably less than the solid phase thickness, one may use the
Dirichlet condition (\ref{38}). It should be stressed that we
neglect the initial distribution of the temperature in the solid
phase and consider the process on the stage when two phases take
already place.

One may claim that formulae (\ref{34})--(\ref{38}) present a class
of BVPs with moving boundaries and take into account  a number of
different  situations, which occur in the melting and evaporation
processes.  Setting $q(t,u) = \mbox{const}, V_1 = \mbox{const}
\Phi(t)$ and $ h(t,u)= \Phi(t)u$, where $\Phi(t)$ is a
correctly-specified function, one obtains the problem, which is the
most typical, see, e.g., \cite{ch93}. In the case of a process when
surfaces are exposed to very  powerful periodic laser pulses these
functions take  complicated forms \cite{ch-od91}.

The BVP obtained is based on the standard nonlinear heat equations.
Lie symmetry of non-coupled system (\ref{34})--(\ref{35}) can be
easily derived using the determining equations from paper
\cite{ch-king4}, where reaction-diffusion systems of more general
form have been investigated. Now we formulate a theorem, which gives
complete information on Lie symmetry of this system.

\begin{theorem}
All possible maximal algebras of invariance (up to equivalent
representations generated by transformations of the form (\ref{39}))
of the system (\ref{34}) and (\ref{35}) for any fixed vectors $(d_1,
d_2)$ with strictly positive functions $d_1(u)$ and $d_2(v)$ are
presented in Table 2. Any other system of the form
(\ref{34}) and (\ref{35}) is reduced to one of those with diffusivities
from Table 2 by an equivalence transformation of the form
\begin{equation}\label{39}
t \rightarrow  e_0t + t_{0},\quad x \rightarrow e_1x + x_{0}, \quad
u \rightarrow  e_2u + u_{0}, \quad v \rightarrow  e_3v + v_{0},
\end{equation}
where $e_i \not=0 \ (i = 0, \ldots,3), t_{0}, x_{0}, u_0$, and
$v_{0}$ are arbitrary parameters.
\end{theorem}

\begin{table}
\caption{Lie algebras of NHEs system (\ref{34})--(\ref{35}).(Here
$k_1,k_2, m$ and $n$ are arbitrary non-zero constants; while
$\alpha(t,x)$ and $\beta(t,x)$ are arbitrary solutions of the linear
heat equations $\alpha_t = k_1 \alpha_{x x}$ and $\beta_t = k_2
\beta_{x x}$, respectively.)}
\renewcommand{\arraystretch}{1.5}
\tabcolsep=10pt
\begin{center}
\begin{tabular}{cccl}
  \hline\hline
  {no} & {$d_1(u)$} & {$d_2(v)$} & {Basic operators of MAI}\\
  \hline
  1. & $\forall$ & $\forall$ & $ A = \langle \partial_{t}, \partial_{x}, 2t \partial_{t} + x \partial_{x}\rangle$\\
  2. & $k_1$ & $\forall$ & $ A, u \partial_u, \alpha(t,x) \partial_u $\\
  3. & $\forall$ & $k_2$ & $ A, v \partial_v, \beta(t,x) \partial_v $\\
  4. & $e^u$ & $e^v$ & $ A, x \partial_{x} + 2 \partial_{u} +  2 \partial_{v}$  \\
  5. & $e^u$ & $v^m$& $ A, x \partial_{x} + 2 \partial_u + \frac{2}{m} v \partial_{v} $ \\
  6. & $u^n$ & $e^v$& $ A, x \partial_{x} + \frac{2}{n} u \partial_u + 2 \partial_{v} $ \\
  7. & $u^n$ & $v^m$& $ A, x \partial_{x} + \frac{2}{n} u \partial_u + \frac{2}{m} v \partial_{v} $ \\
  8. & $u^{-\frac{4}{3}}$ & $v^{-\frac{4}{3}}$ & $ A, x \partial_{x} - \frac{3}{2} u \partial_u - \frac{3}{2} v
  \partial_{v}, x^2 \partial_x - 3 x u \partial _u - 3 x v \partial_v $\\
  9. & $k_1$ & $k_2$ & $A, u \partial_u, v \partial_v, \alpha(t,x) \partial_u, \beta(t,x) \partial_v, 2 t \partial_x - x \left(\frac{1}{k_1} u \partial_u + \frac{1}{k_2}v \partial_v \right),
  $\\ &  &  & $4 t x \partial_x + 4 t^2 \partial_t - \frac{1}{k_1}(x^2 + 2 k_1 t) u \partial_u - \frac{1}{k_2}(x^2 + 2 k_2 t) v \partial_v$ \\
  \hline\hline
\end{tabular}
\end{center}
\end{table}

\begin{remark}
In the case of linear  system
(\ref{34})--(\ref{35}) with $k_1 = k_2$ (see case 9 of Table 2), the
Lie algebra extension occurs by the operators $v
\partial_u$ and $u \partial_v$. However, BVP (\ref{34})--(\ref{38}) with $d_{1}(u)=d_{2}(v)=k_1 $
is rather artificial  from physical point  because diffusivities of
solid and liquid phases must be  different. Thus, we don't consider
this case below.
\end{remark}

\begin{remark}
If one  takes into account the trivial discrete
transformations $t \rightarrow t, \ x \rightarrow x, \ u \rightarrow
v$, and $v \rightarrow u$, then cases 2 and 3, 5 and 6 arising in
Table 2  are equivalent. However,  the class of BVPs
(\ref{34})--(\ref{38}) isn't invariant under these transformations
because of boundary  conditions (\ref{36}) and (\ref{38}). Thus, we
don't take into account them  in what follows.
\end{remark}

Using the set of  equivalence transformations (\ref{39}), we can
straightforwardly extend one to the  relevant set for BVP
(\ref{34})--(\ref{38}) by adding the identical transformations for
the variables $S_k(t,x)$. Direct calculations show that the most
general form of those is
\begin{equation}\label{40}
t \rightarrow  e_0t + t_{0},\ \  x \rightarrow e_1x + x_{0}, \ \  u
\rightarrow  e_2u + u_{0}, \ \  v \rightarrow  e_3v + v_{0}, \ \ S_1
\rightarrow S_1, \ \  S_2 \rightarrow S_2.
\end{equation}
where $e_i \not=0 \ (i = 0, \ldots,3), t_{0}, x_{0}, u_0$, and
$v_{0}$ are arbitrary parameters ($e_1 > 0$).

Now we formulate the main result of this section.

\begin{table}
\caption{Lie invariance of BVP (\ref{34})--(\ref{38})}
\renewcommand{\arraystretch}{1.5}
\tabcolsep=10pt
\begin{center}
\begin{tabular}{cccl}
  \hline\hline
  {no} & {$q(t,u)$} & {$h(t,u)$} & {Lie groups of invariance}\\
  \hline
  1. & $\forall$ & $\forall$ & $t^{\ast} = t, \ x^{\ast} = x + \varepsilon, \ u^{\ast} = u, \ v^{\ast} = v, \ S^{\ast}_1 = S_1, \ S^{\ast}_2 = S_2$\\
  2. & $q(u)$ & $h(u)$ & $t^{\ast} = t + \varepsilon_1, \ x^{\ast} = x + \varepsilon_2, \ u^{\ast} = u, \ v^{\ast} = v, \ S^{\ast}_1 = S_1, \ S^{\ast}_2 = S_2$\\
  3. & $\frac{q(u)}{\sqrt{t}}$ & $\frac{h(u)}{\sqrt{t}}$ & $ t^{\ast} = t e^{2 \varepsilon_1}, \ x^{\ast} = x e^{\varepsilon_1}+ \varepsilon_2, \ u^{\ast} = u, \ v^{\ast} = v, \ S^{\ast}_1 = S_1, \ S^{\ast}_2 = S_2$\\
  \hline\hline
\end{tabular}
\end{center}
\end{table}

\begin{theorem}
BVP (\ref{34})--(\ref{38}) with any smooth functions $d_1(u)$,
$d_2(v), q(t,u), h(t,u)$ and $H_1(u)$ is invariant
 under the one-parameter Lie group presented in
case 1 of Table 3.  All possible extensions of this  Lie group
invariance (up to equivalent representations generated by
equivalence  transformations of the form (\ref{40}))  depend only on
the form of the functions $q(t,u)$ and $ h(t,u)$, and are presented
in cases 2 and 3 of Table 3. Any other BVP of the form
(\ref{34})--(\ref{38}) is invariant under two-parameter Lie group is
reduced by transformations (\ref{40}) to one of those with the
functions $q(t,u)$ and $h(t,u)$ from Table 3.
\end{theorem}

\noindent \textbf{Proof}. According to Definition 2 and Theorem 2
 we need to examine the nine different cases
  listed  in Table 2.
It turns out that the examination of  the first case, when the
functions $d_1(u)$ and $d_2(v)$ are arbitrary, leads to the main
result of the theorem presented in Table 3.

Let us  consider the one-parameter Lie groups corresponding to the
basic operators of algebra $A$. Obviously, BVP
(\ref{34})--(\ref{38}) with arbitrary given  functions is invariant
under  the group of space translations generated by the  operator
$P_x = \partial_x$ and this is listed in the first case of Table 3.
 Since any linear combination $\lambda_1 P_t + \lambda_2
D$  of other two operators is equivalent (up to  transformations
(\ref{40})) either to  $P_t = \partial_t$ (if $\lambda_2 = 0$) or to
$D = 2 t \partial_t + x \partial_x$ (if $\lambda_2 \neq 0$),  we
should separately examine these two operators.

Now we apply  Definition 2 to $D$ . Taking into account that BVP
(\ref{34})--(\ref{38}) has two free boundaries, we construct the
extended Lie group $\widetilde{T}_D$ corresponding to the operator
$D$:
\begin{equation}\label{42}
\widetilde{T}_D :  t^{\ast} = t e^{2 \varepsilon_1}, \ x^{\ast} = x
e^{\varepsilon_1}, \ u^{\ast} = u, \ v^{\ast} = v, \ S^{\ast}_1 =
S_1, \ S^{\ast}_2 = S_2.
\end{equation}
 According to item (c), the boundary
conditions (\ref{36}) are invariant with respect to the group
$\widetilde{T}_D$, if the manifold ${\cal M} = \left \{S_1(t,x) =
0,\ d_{1}(u) \frac{\partial u}{\partial x} = H_1(u) V_1 - q(t,u), \
V_1 - h(t,u) = 0\right \}$ satisfies the conditions
\begin{equation}\label{43}
\left. S_1^{\ast}\right \vert_{\cal M} = 0, \ \left.d_{1}(u^{\ast})
\frac{\partial u^{\ast}}{\partial x^{\ast}} - H_1(u^{\ast})
V^{\ast}_1 + q(t^{\ast},u^{\ast})\right \vert_{\cal M} = 0, \
 \left.V^{\ast}_1 - h(t^{\ast},u^{\ast})\right \vert_{\cal M} = 0.
\end{equation}
\noindent Taking into account (\ref{42}), one finds
\begin{equation}\label{44}
\frac{\partial u^{\ast}}{\partial x^{\ast}} = e^{- \varepsilon_1}
\frac{\partial u}{\partial x}, \ \frac{\partial v^{\ast}}{\partial
x^{\ast}} = e^{- \varepsilon_1} \frac{\partial v}{\partial x}, \
V^{\ast}_k = e^{- \varepsilon_1} V_k, \ k = 1,2,
\end{equation}
\noindent so that the second and third equations of  (\ref{43})
produce the equations
\begin{equation}\label{45}
e^{\varepsilon_1} q\left(t e^{2\varepsilon_1},u \right) = q(t,u)
\quad \ e^{\varepsilon_1} h\left(t e^{2\varepsilon_1},u \right) =
h(t,u),
\end{equation}
 to find  the functions $q(t,u)$ and
$h(t,u)$.  Solving (\ref{45}) one  obtains
\begin{equation}\label{46}
q(t,u) = \frac{q(u)}{\sqrt{t}}, \ \  h(t,u) = \frac{h(u)}{\sqrt{t}},
\end{equation}
where $q(u) $ and $h(u)$ are arbitrary smooth functions. The
invariance criterium of the boundary conditions (\ref{37}) for
$\widetilde{T}_D$ is fulfilled for arbitrary parameters arising in
(\ref{37}), while the invariance of condition
 (\ref{38}) under $T_D$  is obvious. Thus, BVP
(\ref{34})--(\ref{38}) is invariant with respect to the Lie group
$\widetilde{T}_D$ if and only if restrictions (\ref{46})
 take place. This is exactly listed in case 3 of Table 3.

In a quite similar way one can show, that the BVP under study is
invariant with respect to the extended Lie group corresponding to
the operator $P_t = \partial_t$ if and only if the restrictions on
\begin{equation}\label{47}
q(t,u) = q(u) \ \mbox{and} \ h(t,u) = h(u),
\end{equation}
take place, and this is what exactly listed in case 2 of Table 3.

Much more cumbersome calculations are needed to show that there are
no any new Lie group invariance for  BVP (\ref{34})--(\ref{38})
nevertheless there are eight special cases listed in Table 2, which
lead to the extensions of MAI of the basic equations (\ref{34}).

Let us consider case 2 of Table 2. Firstly, we check the invariance
of BVP (\ref{34})--(\ref{38}) with respect to the one-parameter
extended Lie groups corresponding to the operators $X_1 = u
\partial_u$ and $X_2 = \alpha(t,x) \partial_u$:
\begin{equation}\label{d1}
\widetilde{T}_1 :  t^{\ast} = t, \ x^{\ast} = x, \ u^{\ast} = u
e^{\varepsilon_1}, \ v^{\ast} = v, \ S^{\ast}_1 = S_1, \ S^{\ast}_2
= S_2,
\end{equation}
and
\begin{equation}\label{d2}
\widetilde{T}_2 :  t^{\ast} = t, \ x^{\ast} = x, \ u^{\ast} = u +
 \alpha(t,x) \varepsilon_2, \ v^{\ast} = v, \ S^{\ast}_1 = S_1, \
S^{\ast}_2 = S_2.
\end{equation}
According to item (c) of Definition 2, the boundary conditions
(\ref{37}) are invariant with respect to the group
$\widetilde{T}_1$, if the conditions
\begin{equation}\label{d3}
\left. S^{\ast}_2  \right \vert_{\cal N} = 0, \ \left. d_{2}(v_m)
\frac{\partial v^{\ast}}{\partial x^{\ast}} - d_{1}(u_m)
\frac{\partial u^{\ast}}{\partial x^{\ast}} - H_2(v_m) V^{\ast}_2
\right \vert_{\cal N} = 0, \ \left. u^{\ast} - u_m \right
\vert_{\cal N} = 0, \ \left. v^{\ast} - v_m \right \vert_{\cal N} =
0
\end{equation}
are satisfied, where the manifold  \[ {\cal N} = \left \{S_{2}(t,x)
= 0, \ d_{2}(v_m) \frac{\partial v}{\partial x} - d_{1}(u_m)
\frac{\partial u}{\partial x} - H_2(v_m) V_2 = 0, \ u - u_{m} = 0, \
v - v_{m} = 0 \right \}. \] Taking into account (\ref{d1}) and the
second equation of (\ref{d3}),  we arrive at the requirement
\begin{equation}\label{d4}
\frac{\partial u}{\partial x} = \frac{\partial u}{\partial x}
e^{\varepsilon_1} \Rightarrow \varepsilon_1 = 0.
\end{equation}
Similarly, one easily checks   that the boundary conditions
(\ref{37})
 isn't invariant with respect to the Lie group $\widetilde{T}_2$,
too. Indeed, to satisfy  the third equation of (\ref{d3}), one
obtains the requirement
\begin{equation}\label{d5}
 \alpha(t,x) \varepsilon_2 = 0 \Rightarrow \varepsilon_2 = 0.
\end{equation}

Let us now examine the invariance of BVP (\ref{34})--(\ref{38}) with
respect to an extended Lie group $\widetilde{T}_c$ corresponding to
a liner combination of operators $P_t$, $D$, $X_1$, and $X_2$, i.e.
\begin{equation}\label{d6}
X_c = (\lambda_1 + 2 \lambda_2 t) \partial_t + \lambda_2 x
\partial_x + (\lambda_3 u + \lambda_4 \alpha(t,x)) \partial_u,
\end{equation}
where $\lambda_i, \ i =1, \ldots, 4$ are arbitrary parameters and
$\lambda_3^2 + \lambda_4^2 \neq 0$ (otherwise the operator
$\lambda_1 P_t + \lambda_2 D$  is obtained). Having transformations
(\ref{40}),  we can put $\lambda_1 = 0$ and $\lambda_2 = 1$ in
(\ref{d6}) so that the operator  takes the form
\begin{equation}\label{d7}
X_c = 2 t \partial_t + x
\partial_x + (\lambda_3 u + \lambda_4 \alpha(t,x)) \partial_u, \
\lambda_3^2 +  \lambda_4^2  \neq 0.
\end{equation}
The corresponding    Lie group $\widetilde{T}_c$ is
\begin{eqnarray}
& & \widetilde{T}_c :  t^{\ast} = t e^{2 \varepsilon_c}, \ x^{\ast}
= x e^{\varepsilon_c}, \ u^{\ast} = u e^{\lambda_3 \varepsilon_c} +
\lambda_4 \int_0^{\varepsilon_c} \alpha(t e^{2 \tau}, x e^{\tau}) e^
{\lambda_3 (\tau - \varepsilon_c)} d \tau, \ v^{\ast} = v,\nonumber
\\ & & \quad \ \ S^{\ast}_1 = S_1, \ S^{\ast}_2 = S_2.\label{d8}
\end{eqnarray}
Now we again show that  boundary conditions (\ref{37}) are not
invariant under  $\widetilde{T}_c$. In fact, the third equation of
(\ref{d3}) gives  the restriction
\begin{equation}\label{d10}
\lambda_4 \int_0^{\varepsilon_c} \alpha(t e^{2 \tau}, x e^{\tau}) e^
{\lambda_3 (\tau - \varepsilon_c)} d \tau = u_m (1 - e^{\lambda_3
\varepsilon_c}),
\end{equation}
so that  the Lie group $\widetilde{T}_c$  can be written in the form
\begin{equation}\label{d11}
\widetilde{T}_c :  t^{\ast} = t e^{2 \varepsilon_c}, \ x^{\ast} = x
e^{\varepsilon_c}, \ u^{\ast} = u e^{\lambda_3 \varepsilon_c} + u_m
(1 - e^{\lambda_3 \varepsilon_c}), \ v^{\ast} = v, \ S^{\ast}_1 =
S_1, \ S^{\ast}_2 = S_2.
\end{equation}
Taking into account (\ref{d11}) and the second equation of
(\ref{d3}),  we arrive at the requirement
\begin{equation}\label{d12}
\frac{\partial u}{\partial x} = \frac{\partial u}{\partial x}
e^{\lambda_3 \varepsilon_c} \Rightarrow \lambda_3 = 0,
\end{equation}
what leads to $\lambda_4 = 0$ (see (\ref{d10})). However, this
contradicts to the assumption $\lambda_3^2 + \lambda_4^2 \neq 0$.

Hence,  case 2 of Table 2 is completely examined. Cases 3 and 9 of
Table 2  can be studied in a quite similar manner because each of
them  needs to examine  groups (\ref{d1}) and (\ref{d2}).

Consider case 4 from Table 2. Here the   operator $X_4 = x
\partial_x + 2 \partial_u + 2
\partial_v$ arises, which generates   the extended Lie group
$\widetilde{T}_4$:
\begin{equation}\label{d13}
t^{\ast} = t, \ x^{\ast} = x e^{\varepsilon_4}, \ u^{\ast} = u + 2
\varepsilon_4, \ v^{\ast} = v + 2 \varepsilon_4, \  S_1^{\ast} =
S_1, \ S_2^{\ast} = S_2.
\end{equation}
Applying Definition 2 to  the boundary conditions (\ref{37}) in the
case of  $\widetilde{T}_4$,   equations (\ref{d3}) are again
obtained. The third and fourth equations of (\ref{d3}) lead to the
requirement
\begin{equation}\label{d14}
u_m = u_m + 2 \varepsilon_4, \ v_m = v_m + 2 \varepsilon_4,
\end{equation}
hence,  $\varepsilon_4 = 0$. Thus,  BVP (\ref{34})--(\ref{38})
cannot be  invariant with respect to the Lie group
$\widetilde{T}_4$. Moreover, the same result is obtained if one
examines any linear combination of operators $P_t $, $D$, and $X_4$,
i.e.
\begin{equation}\label{d15}
X = (\lambda_1 + 2 \lambda_2 t) \partial_t + (\lambda_2 + \lambda_3)
x \partial_x + 2 \lambda_3 u \partial_u + 2 \lambda_3 v
\partial_v, \ \lambda_3 \neq 0
\end{equation}
with $\lambda_3 \not =0$. Thus, we conclude, that  the exponential
diffusivities arising in  case 4 of Table 2  don't lead to any new
Lie groups of invariance  of  BVP (\ref{34})--(\ref{38}). Cases 5
and 6 of Table 2  can be studied in a quite similar way.

Let us consider case 7 of  Table 2, which needs separate
examination. The extended  Lie group $\widetilde{T}_7$ corresponding
to the operator $X_7 = x
\partial_x + \frac{2}{n} u \partial_u + \frac{2}{m} v \partial_v$
is
\begin{equation}\label{48}
T_7 : t^{\ast} = t, \ x^{\ast} = x e^{\varepsilon_7}, \ u^{\ast} = u
e^{\frac{2}{n}\varepsilon_7}, \ v^{\ast} = v
e^{\frac{2}{m}\varepsilon_7},\ S^{\ast }_1 = S_1, \ S^{ \ast}_2 =
S_2.
\end{equation}

In order to the boundary conditions (\ref{37}) be invariant under
$\widetilde{T}_7$  equations (\ref{d3}) are again obtained. It turns
out, equations (\ref{d3}) are fulfilled if $m>0, n>0, \ u_m=v_m=0$
and $H_2(0)=0$. Hence, we must apply   item (d) of Definition 2 to
the boundary conditions (\ref{38}):
\begin{equation}\label{51}
\left. x^{\ast} \right \vert_{\cal P} = + \infty, \ \left. v^{\ast}
- v_{\infty} \right \vert_{\cal P} = 0,
\end{equation}
where ${\cal P} = \left \{x = + \infty, \ v - v_{\infty} = 0 \right
\}$. The group $\widetilde{T}_7$ transforms the second equation of
(\ref{51}) as follows
\begin{equation}\label{52}
 v_{\infty} e^{\frac{2}{m}\varepsilon_7} - v_{\infty} = 0,
\end{equation}
hence, we arrive at the restriction $ v_{\infty} = 0$ (otherwise
$\varepsilon_7=0$). Thus, the contradiction is obtained because,  in
the very beginning, we assumed
 $v_m \neq v_{\infty}$.
 One may check that the same result is obtained  for  any liner combination of the operators
$P_t$, $D$, and $X_7$ . Thus,  case 7 of Table 2  is completely
examined.

Case 8 of  Table 2  can be treated    in a quite  similar  way as we
did in Theorem 1 (see formulae (\ref{22}) and (\ref{22a})).

The proof is now completed. $\blacksquare$

Finally, we note that theorem 2 from the recent paper
\cite{ch-kov-09} follows as a particular case from Theorem 3 (but
not wise versa !), nevertheless  Definition 2  was not used in
\cite{ch-kov-09}.

\section{\bf Symmetry reduction and invariant solutions of the class BVPs (\ref{34})--(\ref{38})}

In this section we  consider symmetry reduction of  BVPs of the form
(\ref{34})--(\ref{38}) to BVPs for systems of two ordinary
differential equations and construct exact solutions for the reduced
BVPs.

Operator $P_x$, corresponding to the invariance transformations in
case 1 of Table 3,  leads to an ansatz, which  doesn't depend on the
space variable  $x$. This   contradicts to the free boundary
surfaces and  leads to non-physical solutions.

According to  case 2 of Theorem 3, each BVPs belonging to the class
under study for $q(t,u) = q(u)$ and $h(t,u) = h(u)$  admits the
two-dimensional Lie algebra with basic operators $P_t =
\partial_t$ and $P_x = \partial_x$. Hence, it also admits the operator $X_1 =
\partial_t + \mu \partial_x, \ \mu \in \mathbb{R}$, which leads to
the plane-wave ansatz
\begin{equation}\label{4.1}
u = u(\xi), \ v = v(\xi), \ S_k = S_k(\xi), \ \xi = x - \mu t,
\end{equation}
where $k = 1,2$ and $\mu$ is an unknown velocity.  Note, that ansatz
 (\ref{4.1}) with $\mu = 0$ leads to  stationary solutions
of the BVP in question. However, these solutions don't have
essential physical sense so that  will not studied hereafter.

In  case 3 of Table 3, each BVP of the form  (\ref{34})--(\ref{38})
admits two-dimensional Lie algebra with basic operators $P_x$ and $D
= 2 t
\partial_t + x \partial_x$. Obviously, that any linear combination
of these operators is equivalent, up to transformations (\ref{40}),
to operator $D$, which generates the ansatz
\begin{equation}\label{4.2}
u = u(\omega), \ v = v(\omega), \ S_k = S_k(\omega), \ \omega =
\frac{x}{\sqrt t}, \ \ k = 1,2.
\end{equation}

Using  ans\"atze (\ref{4.1})--(\ref{4.2}) and taking into account
the restrictions on the functions $q(t,u)$ and $h(t,u)$ (see Theorem
3) one can easy obtain the following theorems.

\begin{theorem}  Ansatz (\ref{4.1}) reduces
any nonlinear BVP of the form (\ref{34})--(\ref{38}) with the
coefficient restrictions  $q(t,u) = q(u)$ and $h(t,u) = h(u)$  to
the   BVP for the second-order ODEs
\begin{eqnarray}
& &  \frac{d}{d \xi}\left(d_{1}(u)
\frac{d u}{d \xi}\right) + \mu \frac{d u}{d \xi}  = 0, \ \ \ 0 < \xi < \delta,  \label{4.3}\\
 & &  \frac{d}{d \xi}\left(d_{2}(v)
\frac{d v}{d \xi}\right) + \mu \frac{d v}{d \xi}  = 0, \ \ \ \xi > \delta, \label{4.4}\\
& & \qquad \xi = 0: \ d_{1}(u)
\frac{d u}{d \xi} = H_1(u) \mu - q(u), \ \mu = h(u), \label{4.5}\\
& & \qquad \xi = \delta: \ d_{2}(v_m) \frac{d v}{d \xi} = d_{1}(u_m)
\frac{d u}{d \xi} + H_2(v_m) \mu, \ u = u_{m}, \ v =
v_{m}, \label{4.6}\\
& & \qquad \xi  =  +\infty: \ v = v_{0}, \label{4.7}
\end{eqnarray}
where $\delta$ and $\mu$ are to be determined parameters.
\end{theorem}

\begin{theorem}
Ansatz (\ref{4.2}) reduces any nonlinear BVP of the form
(\ref{34})--(\ref{38}) with the coefficient restrictions $q(t,u) =
\frac{q(u)}{\sqrt t}$ and $h(t,u) = \frac{h(u)}{\sqrt t}$ to the BVP
for the second-order ODEs
\begin{eqnarray}
& & \frac{d}{d \omega} \left(d_{1}(u) \frac{d u}{d \omega}\right) +
\frac{\omega}{2} \frac{d u}{d \omega} = 0,
\ \ \ \omega_1 < \omega < \omega_2, \label{4.8}\\
& & \frac{d}{d \omega} \left(d_{2}(v) \frac{d v}{d \omega}\right) +
\frac{\omega}{2} \frac{d v}{d \omega} = 0,
\ \ \ \omega > \omega_2, \label{4.9} \\
&& \qquad \omega = \omega_{1}: d_{1}(u) \frac{d u}{d \omega} =
H_1(u) \frac{\omega_{1}}{2} - q (u), \ \frac{\omega_1}{2} = h(u), \label{4.10}\\
& & \qquad \omega = \omega_{2}: d_{2}(v_m) \frac{d v}{d \omega} =
d_{1}(u_m) \frac{d u}{d \omega} + H_2(v_m) \frac{\omega_{2}}{2},\ u = u_{m}, \ v = v_{m},\label{4.11} \\
& & \qquad \omega = + \infty: v = v_{0},\label{4.12}
\end{eqnarray}
where $\omega_1$ and $\omega_2$ are to be determined parameters.
\end{theorem}

Let us consider BVP (\ref{4.3})--(\ref{4.7}). It is well-known, that
the system of nonlinear ordinary differential equations
(\ref{4.3}) and (\ref{4.4}) with $\mu = 0$ can be linearized by the
Kirchhoff substitution. It turns out, this substitution can be
generalized in the case $\mu \neq 0$ by introducing  new independent
variables\cite{ch93}
\begin{eqnarray}
& & U = u - u_m, \ V = v - v_0, \label{4.13} \\
& & \xi = \int_0^{\eta} d_1(U + u_m) d \eta, \ \ 0 \leq \xi \leq
\delta, \label{4.14} \\
& & \xi = \delta + \int_{\delta^{\ast}}^{\eta} d_2(V + v_0) d \eta,
\ \ \xi \geq \delta, \label{4.15}
\end{eqnarray}
where the lower bonds of integration  are chosen as $\left. \xi
\right \vert_{\eta = 0} = 0$ and $\left. \xi \right \vert_{\eta =
\delta^{\ast}} = \delta$.

Substituting (\ref{4.13})--(\ref{4.15}) into  BVP
(\ref{4.3})--(\ref{4.7}), one obtains the BVP for system of {\it
linear} differential equations
\begin{eqnarray}
& &  \frac{d^2 U}{d \eta^2} + \mu \frac{d U}{d \eta}  = 0, \ \ \ 0 < \eta < \delta^{\ast},  \label{4.16}\\
 & &  \frac{d^2 V}{d \eta^2} + \mu \frac{d V}{d \eta}  = 0, \ \ \ \eta > \delta^{\ast}, \label{4.17}\\
& & \qquad \eta = 0: \
\frac{d U}{d \eta} = H_1(U + u_m) \mu - q(U + u_m), \ U = h^{-1}(\mu) - u_m, \label{4.18}\\
& & \qquad \eta = \delta^{\ast}: \ \frac{d V}{d \eta} = \frac{d U}{d
\eta} + H_2(v_m) \mu, \ U = 0, \ V =
V_{m}, \label{4.19}\\
& & \qquad \eta  =  +\infty: \ V = 0, \label{4.20}
\end{eqnarray}
where $h^{-1}$ is an inverse function to $h(u)$ (hereafter existence of
this function is assumed), $V_m = v_m - v_0$, $\delta^{\ast}$ is new
to be determined parameter.

To solve  BVP (\ref{4.16})--(\ref{4.20}) one needs to find the
unknown functions $U = U(\eta)$ and $V = V(\eta)$ and the parameters
$\mu$ and $\delta^{\ast}$. Since equations
(\ref{4.16}) and (\ref{4.17}) are linear ordinary differential
equations with constant coefficient their general solution is
\begin{eqnarray}
& & U = C_1 + C_2 e^{- \mu \eta}, \ 0 \leq \eta \leq \delta^{\ast}, \label{4.21} \\
& & V = C_3 + C_4 e^{- \mu \eta}, \ \eta \geq \delta^{\ast},
\label{4.22}
\end{eqnarray}
where $C_i (i = 1, \ldots, 4)$ are to be determined constants.

Substituting solution (\ref{4.21}) into the second equation of the
boundary conditions (\ref{4.18}) and the second equation of
(\ref{4.19}), one finds the constants $C_1$ and $C_2$:
\begin{equation}\label{4.23}
C_1 = \left(h^{-1}(\mu) - u_m \right)\frac{e^{- \mu
\delta^{\ast}}}{e^{- \mu \delta^{\ast}} - 1}, \ C_2 =
\left(h^{-1}(\mu) - u_m \right)\frac{1}{e^{- \mu \delta^{\ast}} -
1}.
\end{equation}
Similarly, the constants $C_3$ and $C_4$ can be found using the
third equations from (\ref{4.19}) and the boundary condition
(\ref{4.20})
\begin{equation}\label{4.24}
C_3 = 0, \ C_4 = V_m e^{\mu \delta^{\ast}}.
\end{equation}
Thus,  using formulae (\ref{4.21})--(\ref{4.24}),   we construct the
explicit formulae for the function $U = U(\eta)$ and $V = V(\eta)$:
\begin{eqnarray}
& & U = \left(h^{-1}(\mu) - u_m \right)\frac{e^{\mu ( \delta^{\ast}
-
\eta)} - 1}{e^{ \mu \delta^{\ast}} - 1}, \label{4.25}\\
& & V = V_m e^{ \mu ( \delta^{\ast} - \eta)}. \label{4.26}
\end{eqnarray}
Finally, we  need to specify the parameters $\mu$ and
$\delta^{\ast}$. This can be done by substituting (\ref{4.25}) and
(\ref{4.26}) into the first equations  of the boundary conditions
(\ref{4.18}) and (\ref{4.19}), and using the formulae
\begin{equation}\label{4.27}
\frac{d U}{d \eta} = - \mu \left(h^{-1}(\mu) - u_m
\right)\frac{e^{\mu ( \delta^{\ast} - \eta)} - 1}{e^{ \mu
\delta^{\ast}} - 1}, \ \ \frac{d V}{d \eta} = - \mu V_m e^{ \mu (
\delta^{\ast} - \eta)}.
\end{equation}
After the corresponding calculations, we arrive at the  equations
\begin{eqnarray}
& & \left(h^{-1}(\mu) - u_m \right)\frac{e^{\mu \delta^{\ast}}}{e^{
\mu \delta^{\ast}} - 1} = \frac{q(h^{-1}(\mu))}{\mu} -
H_1(h^{-1}(\mu)),\label{4.28} \\
& & V_m = \left(h^{-1}(\mu) - u_m \right)\frac{1}{e^{ \mu
\delta^{\ast}} - 1} - H_2(v_m).\label{4.29}
\end{eqnarray}
Thus, equation (\ref{4.29}) leads to the explicit formula for the
parameter $\delta^{\ast}$, which corresponds to the thickness of
liquid phase:
\begin{equation}\label{4.30}
\delta^{\ast} = \frac{1}{\mu} \log \left(1 + \frac{h^{-1}(\mu) -
u_m}{V_m + H_2(v_m)} \right).
\end{equation}
Having  (\ref{4.30}),  equations (\ref{4.28}) produces immediately
the transcendent equation to find the velocity $\mu$:\texttt{}
\begin{equation}\label{4.31}
\frac{q(h^{-1}(\mu))}{\mu} - H_1(h^{-1}(\mu)) - h^{-1}(\mu) = V_m -
u_m + H_2(v_m).
\end{equation}

Thus, formulae (\ref{4.25})--(\ref{4.26}) give exact solution of
problem (\ref{4.16})--(\ref{4.20}), where the important parameters
$\delta^{\ast}$ and $\mu$ are defined by expressions (\ref{4.30}) and (\ref{4.31}), respectively. It means that  the exact
solution of BVP (\ref{4.3})--(\ref{4.7})
 is obtained in implicit form. It should be stressed that this result
 is  essential  generalization  of
\cite{ch93}, where a particular case of BVP (\ref{34})--(\ref{38})
was investigated.

\vspace{0.5cm}

\textbf{Example.} We use a model, which was formulated and
investigated by numerical methods in \cite{ch-od91}. The model
describes  the processes of heating, melting and evaporation of
metals under the action at their surface of powerful laser pulses.
Under the relevant assumptions it can be written as follows
\cite{ch-od91}
\begin{eqnarray}
& & \frac{\partial}{\partial
x}\left(\lambda_{1}(T_{1})\frac{\partial T_{1}}{\partial x}\right) =
\rho c_{1}(T_{1}) \frac{\partial
T_{1}}{\partial t}, \ \ 0<s_1(t)<x<s_2(t), \ t>0, \label{4.32} \\
& & \frac{\partial}{\partial x}\left(\lambda_{2}(T_{2})
\frac{\partial T_{2}}{\partial x}\right) = \rho c_{2}(T_{2})
\frac{\partial
T_{2}}{\partial t}, \ \ x>s_2(t), \ t>0, \label{4.33} \\
 & &  \qquad x = s_1(t):\ \lambda_{1}(T_{1})
\frac{\partial T_{1}}{\partial x} = \rho L_v V_1 - Q(t, T_ {1}),
\ V_1 = V_{\ast}\sqrt{\frac{T_v}{T_1}} \ \exp\left(-\frac{T_{\ast}}{T_1}\right), \label{4.34} \\
& &  \qquad x = s_2(t): \ \lambda_{2}(T_{m}) \frac{\partial
T_{2}}{\partial x} = \lambda_{1}(T_{m}) \frac{\partial
T_{1}}{\partial x} + \rho L_m V_2,\ T_{1} = T_{2} = T_{m},
\label{4.35}
\\ & &  \qquad x = +\infty: \ T_{2} = T_{\infty}, \label{4.36}
\end{eqnarray}
where $T_v$, $T_{m}$, $T_{\infty}$ are the known temperatures of
evaporation (under normal atmospheric  pressure), melting and solid
phase of metal, respectively; $\lambda_{k}(T_k)$, $c_{k}(T_k)$,
$\rho$,  $L_v$ and  $L_m$ are the specific heat coefficients
(functions), which are typical for the given metal (note, that we
consider the model with constant and equal densities of solid and
liquid phases of metal, i.e., $\rho_1 = \rho_2 = \rho$); $s_{k}(t)$
are the phase division boundary coordinates to be found; $V_k(t,x) =
\frac{d s_k}{d t}$ are the phase division boundary velocities;
$T_{k}(t, x)$ are the unknown temperature fields; index $k = 1, 2$
corresponds to the liquid and solid phases, respectively.

We will consider the processes of melting and evaporation under a
long time powerful pulse, i.e. $\sim 10^{-3}$ sec and more. In this
case, the function $Q(t,T_1)$  determining the amount of the
absorbed energy  is defined by the  formula
\begin{equation}\label{4.37}
Q(t,T_1) = \chi (T_1)\cdot q_0 (t),
\end{equation}
where $q_0(t)$ is the power of laser pulse, assumed to be a constant
$q_0$, and  $\chi (T_1)$ is the absorbtion coefficient of the
energy. Velocity $V_{\ast}$, which is approximately equal velocity
of sound in the metal, can be given by  the  formula \cite{ch-od91}
\begin{equation}\label{4.38}
V_{\ast} = \frac{P_a \sqrt{A}} {\rho \sqrt{2 \pi R T_v}} \ \exp{
\frac{T_{\ast}}{T_v}}, \ \ T_{\ast} = \frac{A L_v}{R},
\end{equation}
where  $P_a$ is the normal atmospheric pressure, $A$ is the atomic weight,
$R$ is the universal gas constant.

Since our aim is to demonstrate that the formulae obtained above
produce realistic data, we need to specify all coefficients arising
in BVP (\ref{4.32})--(\ref{4.36}). Thus, all coefficients were taken
from the paper \cite{ch-od91}, where   the processes of melting and
evaporation of aluminium were studied. In the particular case, the
temperature dependence of the quantities $c_k(T_k)$ and $\chi(T_1)$
are essential and have the  form
\begin{equation}\label{4.39}
c_1(T_1) = 1086, \ T \geq T_m, \quad
 c_2(T_2) = 752.2 + 0.473 \cdot T, \ T < T_m,
\end{equation}
\begin{equation}\label{4.40}
\chi(T_1) = 0.64 \left(\frac{T_1}{11600}\right)^{0.4},
\end{equation}
where $[c_k(T_k)] = \mbox{J} \mbox{kg}^{-1} \mbox{K}^{-1}$. Other
physical quantities are assumed to be some constants, namely:
$\lambda_1 = \lambda_2 = 240 \ \mbox{W} \mbox{m}^{-1}
\mbox{K}^{-1}$, $\rho = 2545 \ \mbox{kg} \mbox{m}^{-3}$, $L_m = 0.64
\cdot 10^6 \ \mbox{J}\mbox{kg}^{-1}$, $L_v = 10.8 \cdot 10^6 \
\mbox{J}\mbox{kg}^{-1}$, $T_v = 2793 \ \mbox{K}$, $T_m = 933 \
\mbox{K}$, $T_{\infty} = 300 \ \mbox{K}$.

First of all,  to use the results obtained above, we should
transform the governing equations of  BVP (\ref{4.32})--(\ref{4.36})
to the form  (\ref{34}) and (\ref{35}). Using Goodman's substitution
\begin{equation}\label{4.41}
u = \phi_1(T_1) \equiv \int\limits_0^{T_{1}} {c_{1}(\zeta)
\rho}\,d\zeta, \quad v = \phi_2(T_2) \equiv \int\limits_0^{T_{2}}
{c_{2}(\xi) \rho}\,d\xi.
\end{equation}
one easy transforms  BVP (\ref{4.32})--(\ref{4.36}) to the form
(\ref{34})--(\ref{38}), where
\begin{equation}\label{4.42}
d_1(u) = \frac{\lambda_1}{\rho c_1}, \ \ d_2(v) =
\frac{\lambda_2}{\rho} \cdot \frac{1}{\sqrt{a_2^2 + 2 \frac{b_2}{\rho} v}},
\end{equation}
\begin{equation}\label{4.43}
q(u) = \frac{0.64}{11600^{0.4}} \cdot \frac{q_0}{(\rho c_1)^{0.4}} \
u^{0.4}, \ \ h(u) = \frac{P_a \sqrt{c_1 A}}{\sqrt{2 \pi R \rho}} \
\exp{\frac{T_{\ast}}{T_v}} \cdot \frac{1}{\sqrt{u}} \ \exp\left(-
\frac{\rho c_1 T_{\ast}}{u}\right),
\end{equation}
\begin{equation}\label{4.45}
u_m = \rho c_1 T_m, \ \ v_m = \rho \left(a_2 T_m +
\frac{b_2}{2}T_m^2 \right), \ \ v_{\infty} = \rho \left(a_2
T_{\infty} + \frac{b_2}{2}T_{\infty}^2 \right),
\end{equation}
and
\begin{equation}\label{4.46}
H_1 = \rho L_v, \ \ H_2 = \rho L_m.
\end{equation}
In  formulae (\ref{4.42})--(\ref{4.46}), we  use the notation
$c_2(T_2) = a_2 + b_2 T_2$, where the coefficients $a_2$ and $b_2$
are determined by (\ref{4.39}).

Now, using formulae (\ref{4.25}) and (\ref{4.26}), and
(\ref{4.30})and (\ref{4.31}), and making the relevant simplifications,
one can receive the  exact solution of  BVP
(\ref{4.32})--(\ref{4.36}) in  the explicit form
\begin{equation}\label{4.47}
T_1 = T_m + \frac{\left(K + h^{-1}(\mu) - \rho c_1 T_m \right) e^{-
\frac{\mu \rho c_1}{\lambda_1} \xi} - K}{\rho c_1}
\end{equation}
\begin{equation}\label{4.48}
T_2 = \frac{T_{\infty} c_2 \left(\frac{T_m + T_{\infty}}{2} \right)
+ (T_m - T_{\infty}) c_2 \left(\frac{T_{\infty}}{2} \right) e^{-
\frac{\mu \rho c_2(T_{\infty})}{\lambda_2} (\xi - \delta)}}{c_2
\left(\frac{T_m + T_{\infty}}{2} \right) - \frac{b_2}{2} (T_m -
T_{\infty}) e^{- \frac{\mu \rho c_2(T_{\infty})}{\lambda_2} (\xi -
\delta)}}
\end{equation}
\begin{equation}\label{4.49}
\delta = \frac{\lambda_1}{\mu \rho c_1} \log \left(1 +
\frac{h^{-1}(\mu) - \rho c_1 T_m}{K} \right),
\end{equation}
where $K = (T_m - T_{\infty}) \rho c_2 \left(\frac{T_m +
T_{\infty}}{2} \right) + H_2$  and the velocity  $\mu$   satisfies
the transcendent equation
\begin{equation}\label{4.50}
\frac{q(h^{-1}(\mu))}{\mu} - h^{-1}(\mu) = K - \rho c_1 T_m + H_1.
\end{equation}

Equation (\ref{4.50}) can be easy solved numerically  by means of
Maple (Mathematica etc.) program package. We  used  Maple 12. The
calculations were carried out for two values of the parameter $q_0$:
1) $q_0 = 1 \cdot 10^{10} \ \mbox{W} \mbox{m}^{-2}$; 2) $q_0 = 5
\cdot 10^{10} \ \mbox{W} \mbox{m}^{-2}$. The following phase
division boundary velocities ($[\mu]= \mbox{m} \cdot \mbox{c}^{-1}
$) were obtained:
\[
\mu = \left \{\begin{array}{l} 0.10 \ \ \mbox{if} \ \ q_0 = 1 \cdot
10^{10}, \\ 0.54 \ \ \mbox{if} \ \ q_0 = 5 \cdot 10^{10}.
\end{array} \right.
\]
The corespondent liquid phase thickness ($[\delta]= \mbox{m}$) is
\[
s_2(t)-s_1(t)=\delta = \left \{\begin{array}{l} 9.60 \cdot 10^{-4} \
\ \mbox{if} \ \ q_0 = 1 \cdot 10^{10}, \\ 2.23 \cdot 10^{-4} \ \
\mbox{if} \ \ q_0 = 5 \cdot 10^{10}.
\end{array} \right.
\]
The temperature fields of liquid and solid phases of aluminium are
presented in Fig.1. Comparing the temperature fields and the liquid
phase thickness obtained here with those from \cite{ch-od91}, one
concludes that they are sufficiently similar. Of course, one should
take into account that a chain of laser pulses was used in
\cite{ch-od91} for numerical simulations, while one only laser pulse
was used to obtain formulae presented above.

\begin{figure}
\begin{center}
\vspace{1cm}
\includegraphics[width=7 cm]{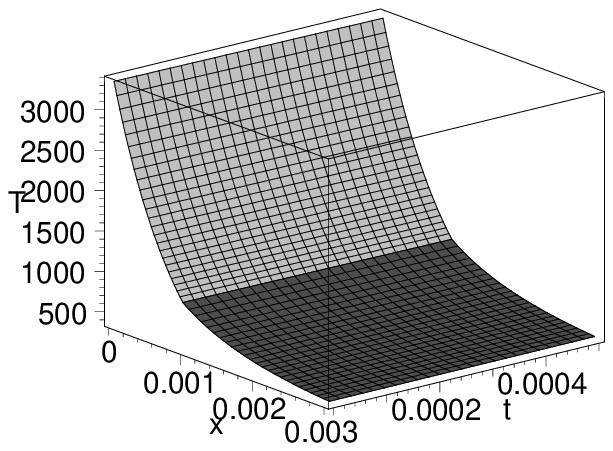}
\includegraphics[width=7 cm]{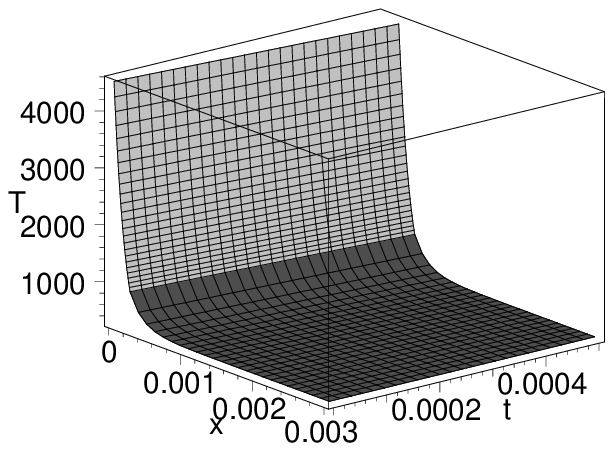}
\caption{Exact solutions of problem (\ref{4.32})--(\ref{4.36}) for
aluminium with the energy flux: 1) $q_0 = 1.0 \cdot 10^{10} \
\mbox{W} \mbox{m}^{-2}$ and 2) $q_0 = 5.0 \cdot 10^{10} \ \mbox{W}
\mbox{m}^{-2}$.}
\end{center}
\end{figure}

\section{\bf Conclusions}

In this paper, nonlinear boundary value problems by means of
the classical Lie symmetry method are studied. First of all, an
analysis of  the known  definitions  of Lie invariance
  for BVPs  is presented.
  Having this done, a new
definition of invariance in Lie sense for  BVPs of the form
(\ref{1})--(\ref{4}) is formulated. This definition (see Definition
2) is applicable to a wide class of BVPs, including those with
several basic equations, with moving boundaries, and with boundary
conditions on non-regular manifolds, moreover, it can be easily
generalized on BVPs with hyperbolic and elliptic basic equations.

The class of  two-dimensional nonlinear BVPs (\ref{34})--(\ref{38}),
modeling the process of melting and evaporation of metals (under
acting  a powerful flux of energy) is studied
 in details.
 Using Definition 2, all possible Lie
symmetries (see Theorem 3) and the relevant reductions with physical
meaning to BVPs for ordinary differential equations (see Theorems 4
and 5) are constructed. The example how to construct exact solution
of the nonlinear problem (\ref{34})--(\ref{38}) with
correctly-specified coefficients for aluminium  is also presented.
We established that {\it exact formulae }
(\ref{4.47})--(\ref{4.49}), obtained by direct application of
Theorem 4, lead to quite realistic results, which are sufficiently
similar to those obtained earlier by {\it numerical simulations}.

The work is in progress to extend the results on  {\it
multidimensional BVPs} using the definition proposed in this paper.

\end{document}